\documentclass[12pt,epsfig,color]{article}
\usepackage{amsmath}
\usepackage{epsfig}
\usepackage{amssymb}
\usepackage{cite}
\input{epsf}
\def\beq{\begin{equation}}
\def\eeq{\end{equation}}
\def\beqa{\begin{eqnarray}}
\def\eeqa{\end{eqnarray}}

\newcommand\as{\alpha_{\mathrm{S}}}
\newcommand\f[2]{\frac{#1}{#2}}

\def\beq{\begin{equation}}
\def\eeq{\end{equation}}
\def\beeq{\begin{eqnarray}}
\def\eeeq{\end{eqnarray}}
\def\to{\rightarrow}
\def\nn{\nonumber}

\def\b0{b_0}

\begin{document}

\begin{titlepage}
\renewcommand{\thefootnote}{\fnsymbol{footnote}}
\begin{flushright}
     \end{flushright}
\par \vspace{10mm}
\begin{center}
{\large \bf
Single-inclusive jet production in polarized $pp$ collisions \\
  at RHIC to next-to-leading logarithmic accuracy}

\end{center}
\par \vspace{2mm}
\begin{center}
{\bf Daniel de Florian}
\hskip .2cm
and
\hskip .2cm
{\bf Federico Wagner  }\\
\vspace{5mm}
Departamento de F\'\i sica, FCEyN, Universidad de Buenos Aires,\\
(1428) Pabell\'on 1 Ciudad Universitaria, Capital Federal, Argentina\\

\end{center}


\par \vspace{9mm}
\begin{center} {\large \bf Abstract} \end{center}
\begin{quote}
\pretolerance 10000

We perform the resummation of large logarithmic corrections to the partonic cross sections for single-inclusive jet production in polarized $pp$ collisions. 
We reach the next-to-leading logarithmic accuracy for this observable with the corresponding matching to the next-to-leading order calculation performed in the small-cone approximation. We present numerical results for the BNL-RHIC collider at $\sqrt{S}=200$~GeV and at $\sqrt{S}=500$~GeV. We find an enhancement of the spin-dependent cross section, specially at high transverse momentum for the jet, resulting in a rather small increase of the double-spin asymmetry $A^{jet}_{LL}$ for this process. 
\end{quote}

\end{titlepage}

\setcounter{footnote}{1}
\renewcommand{\thefootnote}{\fnsymbol{footnote}}


\section{Introduction}

The spin structure of the nucleon continues to be a particular focus of
modern nuclear and particle physics. 
As is well known, the total quark and 
anti-quark spin contribution to the nucleon spin 
was found to be only about $\sim 25 \%$, implying that the gluon spin 
contribution and/or orbital angular momenta may play an important role. 
There is currently much experimental activity aiming at further unraveling 
the nucleon's spin structure.
The main goal of the spin program, besides obtaining the partonic share to the total 
spin of the nucleon, 
 is the extraction of  
 the full set of the $x-$dependent polarized quark ($\Delta q = q^{\uparrow}-q^{\downarrow}$) 
and gluon ($\Delta g=g^{\uparrow}-g^{\downarrow}$) densities of the nucleon. 
Many phenomenological analyses \cite{reviewth} 
demonstrate, however, that available DIS data alone are not sufficient for this purpose.
This is true in particular for $\Delta g(x,Q^2)$ since it contributes to 
DIS in leading-order (LO) only via the $Q^2$ dependence of the polarized structure function $g_1$,  which could not yet be 
accurately studied experimentally. As a result of this, it turns out 
 that the $x$ shape  of $\Delta g$ seems to be hardly 
constrained by the DIS data.

The precise extraction of $\Delta g$ thus remains one of the most interesting 
challenges for spin physics experiments. 
The RHIC collider at BNL, running in 
a  proton--proton mode with longitudinally polarized beams provides
the ideal tool for that purpose. The observables measured so far include 
single-pion production at center-of-mass energy $\sqrt{s}=62$ \cite{Adare:2008qb} and 200 GeV \cite{Adare:2007dg} 
and jet production at  $\sqrt{s}=200$ GeV \cite{Abelev:2007vt}. Unlike DIS, those processes
have a direct gluonic contribution already at the lowest order.

A next-to-leading order (NLO) global analysis that includes all available data 
from inclusive and semi-inclusive polarized deep-inelastic scattering, as well as 
from polarized proton--proton scattering at RHIC
has been recently performed \cite{dssv}. The main outcome of the analysis is the indication of
a rather small gluon polarization in the nucleon over the limited region of momentum fraction
$0.05\lesssim x \lesssim 0.2$. As shown in Ref. \cite{dssv}, single-inclusive jet production plays a very important role towards the determination of the polarized gluon density in that kinematical region. 

While the general description of  single-jet production at {\it unpolarized} hadronic colliders using next-to-leading order (NLO) calculations is rather satisfactory, is has been shown that higher order contributions originated from threshold effects are non-negligible \cite{KO,resjetunp}. At partonic threshold the phase space for gluon bremsstrahlung vanishes, so that only soft and collinear emission is allowed resulting in large logarithmic corrections to the partonic cross section. If we consider the cross section as a function of the jet transverse momentum $p_T$, integrated over the jet rapidity, the large contributions near threshold (reached when $\sqrt{s}=2 p_T$, with $s$ the partonic center-of-mass energy) arise as
$\alpha_S^k(p_T) \log^{2m}(1-\hat{x}_T^2)$ at the $k$th order in perturbation theory, where $m\le k$ and $\hat{x}_T=2 p_T/\sqrt{s}$. 

Even if $p_T$ is large so that $\as(p_T)$ is small, sufficiently
close to threshold the logarithmic terms will spoil the perturbative
expansion to any fixed order. Threshold 
resummation ~\cite{dyresum,Catani:1996yz,KS,KOS,KOS1,LOS,BCMN},
however, allows to reinstate a useful perturbative series by 
systematically taking into account the terms $\as^k \ln^{2m}
\left(1-\hat{x}_T^2\right)$ to all orders in $\as$. This is achieved
after taking a Mellin transform of the hadronic cross section in 
$x_T=2 p_T/\sqrt{S}$, with $\sqrt{S}$ the hadronic c.m. energy. The 
threshold logarithms exponentiate in transform space. In line with the work presented in \cite{resjetunp} we rely on the ``small-cone approximation'' (SCA) \cite{Sterman:1977wj} for the single-inclusive jet cross section in order to be able to perform the resummation in a fully analytical way.

Concerning the phenomenological impact of the resummation, the analysis performed in Ref.\cite{resjetunp} shows
that for the unpolarized cross section terms originated from soft-gluon radiation can alter the fixed order prediction at
RHIC by up to 10$\%$. While the uncertainties on the polarized gluon distribution used to be large enough to generate
differences in the hadronic cross sections much larger than that effect, the situation has drastically change since the
advent of RHIC data. As shown in Ref.\cite{dssv}, not only the gluon polarization in the kinematical region relevant for RHIC
turns out to be rather small but also the corresponding uncertainty is considerably reduced, such that a ${\cal O}(10\%)$
effect might become relevant. Therefore, it is worth to study whether a similar correction arises in the polarized cross
section originated from soft gluon emission. More specifically, we want to know to what extent the polarized asymmetry
is modified by threshold resummation and how that might affect the extraction of the polarized gluon distribution from the experimental data.

The outline of this paper is as follows: In Section~\ref{sec2} after some definitions and notation, we present the expressions for the cross sections at next-to leading logarithmic accuracy (NLL), i.e. including the next-to dominant threshold logarithmic terms to all orders in $\as$, for single-inclusive jet production. In Section~\ref{sec4} phenomenological results are presented at $\sqrt{S}=200$~GeV and at $\sqrt{S}=500$~GeV for the BNL-RHIC collider. In particular we show the cross sections, the corresponding predictions for the asymmetry $A_{LL}^{jet}$ and study the scale dependence of the process. We summarize the main results in Sec.~\ref{sec5}. Finally we compile some necessary results in the appendix. 
  

\section{Single-inclusive jet cross section in perturbation theory \label{sec2}}

We consider single-jet production in high-energy proton--proton collisions when both protons are longitudinally polarized, i.e. $ p(p_1, \Lambda_1) + p(p_2, \Lambda_2) \rightarrow \textrm{Jet($p_{J}$)} + X $ where the $p_i$ ($i=1,2$) represent the four-momenta of the initial protons and $p_J$ the four-momenta of the jet. $\Lambda_i$ denotes the helicity of the initial protons. 

Our main aim is to calculate the spin cross section for this process to NLL accuracy and to evaluate the double-spin asymmetry $A_{LL}^{jet}$. It is customary to define the spin-averaged and spin-dependent cross sections as
\begin{eqnarray}
d \sigma &=& \frac{1}{2}\left[ 
d \sigma(\Lambda_1= +, \Lambda_2= +) + d \sigma(\Lambda_1= +, \Lambda_2= -)\right]\equiv d \sigma^{++} + d \sigma^{+-} \, \nonumber \\
d \Delta \sigma &=& \frac{1}{2}\left[
d \sigma(\Lambda_1= +, \Lambda_2= +) - d \sigma(\Lambda_1= +, \Lambda_2= -)\right]\equiv d \sigma^{++} - d \sigma^{+-}
\end{eqnarray}
respectively, and the corresponding double-spin asymmetry as
\beq
A_{LL}^{jet}= \frac{d \Delta \sigma}{d\sigma} \; . \label{alldef}
\eeq
 
 In this work the four-momenta of the jet is defined as the sum of the four-momenta of all particles within a cone of aperture $R$ in pseudorapidity and azimuthal angle around the jet axis. 
In order to be able to perform the calculation in an analytical form we rely on the ``small-cone approximation'' (SCA). The SCA assumes that the jet cone is rather narrow  and the partonic sections can be computed as an expansion around $\delta \equiv R/ cosh \, \eta$, where $\eta$ is the jet's pseudorapidity. At small $\delta$ the behavior of the jet cross section is of the form $\mathcal{A} \log(\delta) + \mathcal{B} + {\mathcal O}(\delta^2)$. The SCA remains a very good approximation even for relatively large cone sizes of up to $R \simeq 0.7$~\cite{nlojet,resjetunp}.

The jet is produced at relatively large transverse momentum $p_T$. Therefore for such a large-momentum-transfer
reaction, the factorization theorem~\cite{FT} allows one to express the spin-dependent cross section for 
$pp\to \rm{Jet}\, X$  in terms of convolutions of parton
distribution functions for the initial hadrons, and partonic hard-scattering functions. After integration over the pseudorapidity of the jet the transverse momentum distribution reads~\footnote{It is worth noting that dropping all $\Delta$'s in Eq.~(\ref{eq:1}) the usual unpolarized scattering cross section is recovered.}
\begin{align} \label{eq:1}
\frac{p_T^3 d \Delta \sigma^{SCA} (x_T)}{d p_T} = \sum_{a,b}\, \int dx_1 dx_2 & \, \Delta f_{a/H_1}\left(x_1,\mu_F^2\right) 
\Delta f_{b/H_2} \left(x_2,\mu_F^2\right)  \int_0^1
d\hat{x}_T \, \, \delta\left(\hat{x}_T-\f{x_T}{z\sqrt{x_1
x_2}}\right) \,  \nn \\ &\times
\int_{\hat{\eta}_{-}}^{\hat{\eta}_{+}} d\hat{\eta}
\, \f{\hat{x}_T^4 \, s}{2} \,
\f{d \Delta \hat{\sigma}_{ab}\left(\hat{x}_T^2 , \hat{\eta}, R \right)}{d\hat{x}_T^2 d\hat{\eta}}\, ,
\end{align}
where the $\Delta f_{a,b}$ are the spin-dependent parton distributions
of the corresponding hadron, 
\beq
\Delta f_a(x,\mu_F^2)=f_a^+(x,\mu_F^2)-f_a^-(x,\mu_F^2)
\eeq
with $f_a^+$ ($f_a^-$) denoting the distribution of parton type $a$
with positive (negative) helicity in a hadron of positive helicity.
The sum in Eq.~(\ref{eq:1})
runs over all partonic channels, with the associated
spin-dependent partonic cross sections $d\Delta 
\hat{\sigma}_{ab\rightarrow Jet X}$. The relevant subprocesses here are those initiated by $q q', \, q \bar{q}', \, q q, \, q \bar{q}, \, qg$, and $gg$. The dependence of the partonic cross section on the factorization $\mu_F$ and renormalization $\mu_R$ scales is implicit in Eq.(\ref{eq:1}). Along the rest of this paper we fix both scales to be equal and denoted by $\mu\equiv \mu_F=\mu_R$.
Finally, $\hat{\eta}$ is the jet's pseudorapidity at 
parton level, related to the one at hadron level by $\hat{\eta}=\eta 
-\frac{1}{2}\ln(x_1/x_2)$. Its limits are given by
$\hat{\eta}_{+}=-\hat{\eta}_{-}=\ln\left[(1+\sqrt{1-\hat{x}_T^2})/
\hat{x}_T\right]$ where
$s=x_1x_2 S$.
 As discussed in Ref.~\cite{DW1}, a major simplification of the resummation 
formalism occurs when the cross section is integrated over all pseudorapidities $\eta$. For that reason, the results obtained along this paper formally apply only for the transverse momentum inclusive distribution. Nevertheless, the effects of the resummation can be safely applied to the fully differential cross section at central rapidities by simply rescaling the resummed result with the appropriate ratio of NLO cross sections \cite{stermanvogelsang}.

\subsection{Resummed cross section}

To carry out the threshold resummation of the soft-gluon contributions it is customary to take the Mellin transform of the rapidity-integrated cross section in the scaling variable $x_T^2$:
\begin{align}
\label{eq:moments}
\Delta \sigma(N)\equiv \int_0^1 dx_T^2 \, \left(x_T^2 \right)^{N-1} \;
\f{p_T^3\, d\Delta \sigma(x_T)}{dp_T} \, .
\end{align}

In Mellin-moment space, the convolutions in 
Eq.~(\ref{eq:1}) between parton distributions and subprocess cross sections become ordinary products and the threshold logarithms turn into  logarithms in the moment variable $N$. The leading logarithms are of the form
$\as^k \ln^{2k}N$; subleading ones are down by one or more powers
of $\ln N$. Threshold resummation results in exponentiation of 
the soft-gluon corrections in moment space~\cite{dyresum,KOS}. 
The leading logarithms are contained in radiative factors for the 
initial and final partons. Because of color interferences and 
correlations in large-angle soft-gluon emission at NLL, for QCD
hard scattering the resummed cross section becomes a sum of 
exponentials.

Combining results of~\cite{resjetunp,dyresum,Catani:1996yz,KS,KOS,KOS1,LOS,BCMN,dwf}, we can cast the resummed 
spin-dependent partonic cross section for each subprocess 
into a relatively simple form~\footnote{Note that the 
symbols ${\cal D}_N^i, {\cal D}^{{\rm (int)} ab\rightarrow cd}_{I\, N}$ 
in the equation below are usually referred to as 
$\Delta_N^i,\Delta^{{\rm (int)} ab\rightarrow cd}_{I\, N}$ 
in the literature~\cite{DW1}. We have changed this notation
in order to avoid confusion with the label ``$\Delta$'' indicating
spin-dependent cross sections and parton distributions in this paper.}:
\begin{align}
\label{eq:res}
\Delta \hat{\sigma}^{{\rm (res)}}_{ab} (N)= \sum_{c,d}\Delta C_{ab}\,
{\cal D}^a_N\, {\cal D}^{b}_N\, {J'}^{c}_N\,
J^{d}_N\, \left[ \sum_{I} \Delta G^{I}_{ab\to cd}\,
{\cal D}^{{\rm (int)} ab\rightarrow cd}_{I\, N}\right] \,
\Delta \hat{\sigma}^{{\rm (Born)}}_{ab\to cd} (N) \;  ,
\end{align}
here $c$ and $d$ are the possible partons in the final state, whereas the $I$ index takes into account the color configurations of the hard scattering.  $\Delta \hat{\sigma}^{{\rm (Born)}}_{ab\to cd}(N)$ are the same as for hadron production and can be found in Ref.\cite{dwf}. As we mentioned above each of the functions  
$J^{i}_N$,${\cal D}^{i}_N$,${\cal D}^{{\rm (int)} ab\rightarrow cd}_{I\, N}$
in Eq.~(\ref{eq:res}) is given by an exponential. ${\cal D}^a_N$ represents the 
effects of soft-gluon radiation collinear to initial parton $a$ and is
given, in the $\overline{{\mathrm{MS}}}$ scheme, by
\begin{align}\label{Dfct}
\ln {\cal D}^a_N&=  \int_0^1 \f{z^{N-1}-1}{1-z}
\int_{\mu^2}^{(1-z)^2 Q^2} \f{dq^2}{q^2} A_a(\as(q^2)) dz\; ,
\end{align}
and similarly for ${\cal D}^b_N$. Here, $Q^2=2 p_T^2$. We will 
specify the function $A_a$ below. The function $J^{d}_N$ embodies collinear, soft or hard, emission 
by the non-observed recoiling parton $d$ and reads:
\begin{align} \label{Jfct}
\ln J^d_N&=  \int_0^1 \f{z^{N-1}-1}{1-z} \Big[
\int_{(1-z)^2 Q^2}^{(1-z) Q^2} \f{dq^2}{q^2} A_d(\as(q^2)) +
\f{1}{2} B_d(\as((1-z)Q^2)) \Big] dz
\end{align}
whereas the {\it massive} observed jet is accounted by~\cite{resjetunp}:
\begin{equation} \label{Jfctprim}
\ln J^{' c}_N=  \int_0^1 \f{z^{N-1}-1}{1-z} C^{'}_{c} (\as((1-z)^2 Q^2)) dz .
\end{equation}
The coefficients $C^{'(1)}_{c}$ needed in the case of massive jet at threshold are obtained by comparing the first order expansion of the resummed formula to the analytic NLO results in the SCA. They are universal and read:
\begin{equation} \label{Jfctprim2}
 C^{'(1)}_{c}= - C_a \log(\frac{\delta^2}{8}), 
\end{equation}
with $C_g=N_c$ and $C_q=(N_c^2-1)/2N_c$. 
This coefficient contains the dependence on $\log (\delta)$ that regularizes the final-state collinear configuration. Exponentiation in Eq.(\ref{Jfctprim}) provides one power of $\log (\delta)$ for each perturbative order.

Large-angle soft-gluon emission is described by ${\cal D}^{{\rm (int)} ab\rightarrow cd}_{I\, N}$, which depends on
the color configuration $I$ of the participating partons. The sum over color in Eq.~(\ref{eq:res}) is weighted for each color configuration with $\Delta G^{I}_{ab\to cd}$, satisfying $\sum_I \Delta G^{I}_{ab\to cd}=1$. Each of the 
${\cal D}^{{\rm (int)} ab\rightarrow cd}_{I\, N}$ is given by
\begin{align}\label{Dintfct}
\ln{\cal D}^{{\rm (int)} ab\rightarrow cd}_{I\, N} &=
 \int_0^1 \f{z^{N-1}-1}{1-z} D_{I\, ab\to cd}(\as((1-z)^2 Q^2))dz \; .
\end{align}
Finally, the coefficients $\Delta C_{ab}$ contain
$N-$independent hard contributions arising from one-loop
virtual corrections and non-logarithmic soft contributions.

The functions $A_a$, $B_a$, and $D_{I\, ab\to cd}$ in Eqs.~(\ref{Dfct})-(\ref{Dintfct}) are associated with either soft-gluon  or  final-state collinear emission, both independent on the polarization of the initial state partons. Therefore, their expressions are exactly the same as those computed for spin-averaged scattering. Each of these functions ${\cal F}\equiv A_a$, $B_a$, $D_{I\, ab\to cd}$ is a perturbative series in $\as$,
\begin{equation}
{\cal F}(\as)=\frac{\as}{\pi} {\cal F}^{(1)} +
\left( \frac{\as}{\pi}\right)^2 {\cal F}^{(2)} + \ldots \; ,
\end{equation}
where the coefficients $A_a^{(1)}, A_a^{(2)}, B_a^{(1)}$ and $D_{I\, ab \to c d}^{(1)}$ needed to reach next-to-leading logarithmic accuracy can be found in Refs.~\cite{dwf,DW1,resjetunp}.

The only differences between the spin-dependent and the spin-averaged
cases reside on the coefficients $\Delta G^{I}_{ab\to cd},
\Delta C_{ab}$ and the Born cross sections, representing contributions related to hard radiation, which depends on the polarization of the initial state.  Nevertheless both $\Delta \hat{\sigma}^{{\rm (Born)}}_{ab\to cd}(N)$ and  $\Delta G^{I}_{ab\to cd}$ only depend on the corresponding $2\to 2$ partonic subprocess $ab\to cd$, and therefore are the same for jet or hadron production. We implement in Eq.~(\ref{eq:res}) the corresponding results from our previous calculation in \cite{dwf}. 

Therefore, only the coefficients $\Delta C_{ab}$ need to be computed for $pp \to {\rm Jet} \, X$, and their perturbative expansion reads:
\begin{eqnarray}
\Delta C_{ab} = 1 + \frac{\as}{\pi} \Delta C_{ab}^{(1)} + 
{\cal O}(\as^2) \; .
\end{eqnarray}
In order to determine the coefficients $\Delta C_{ab}^{(1)} $, we take
advantage of the full analytic NLO calculation of Ref.~\cite{nlojet}. For
each partonic channel one expands the resummed cross section in 
Eq.~(\ref{eq:res}) to first order in $\as$. Near threshold, one
can straightforwardly take Mellin moments of the full NLO expressions
of~\cite{nlojet}. By comparison of the two results one first verifies
that all logarithmic terms in the full NLO results are correctly 
reproduced by the resummation formalism. The remaining $N$-independent
terms in the NLO cross section give the coefficients 
$\Delta C_{ab\to cd}^{(1)}$. These turn out to have rather lengthy 
expressions, and we only provide them in numerical form in the Appendix.

This completes the collection of the ingredients needed for the resummed
partonic cross section. In the exponents, the large logarithms in 
$N$ occur only as {\it single} logarithms, of the form
$\as^k \ln^{k+1}(N)$ for the leading terms. Next-to-leading logarithms
are of the form $\as^k \ln^k(N)$. Knowledge of the
coefficients given above allows to resum the full LL and NLL full towers
in the exponent. It is useful to expand the resummed exponents 
to definite logarithmic order:
\beeq \label{lndeltams} \!\!\! \!\!\! \!\!\! \!\!\!
\!\!\! \ln {\cal D}_N^a(\as(\mu^2),Q^2/\mu^2)
&\!\!=\!\!& \ln N \;h_a^{(1)}(\lambda) +
h_a^{(2)}(\lambda,Q^2/\mu^2) + {\cal O}\left(\as(\as
\ln N)^k\right) \,,\\ \label{lnjfun} \ln
J_N^a(\as(\mu^2),Q^2/\mu^2) &\!\!=\!\!& \ln N \;
f_a^{(1)}(\lambda) + f_a^{(2)}(\lambda,Q^2/\mu^2) + {\cal
O}\left(\as(\as \ln N)^k\right) \; , \\
\label{masslnj} \ln
J_N^{'a}(\as(\mu^2)) &\!\!=\!\!& \frac{C_a^{'(1)}}{2 \pi b_0} \;\ln (1-2\lambda) + {\cal O}\left(\as(\as \ln
N)^k\right) \, ,
 \eeeq 
where $\lambda=\b0 \as(\mu^2) \ln N$. The functions $h^{(i)}$ and
$f^{(i)}$ are available in~\cite{DW1}. $h^{(1)}$ and $f^{(1)}$ 
contain all LL terms in the perturbative series, while $h^{(2)}$ and $f^{(2)}$
are only of NLL accuracy. Finally, for a complete NLL resummation one also
needs the coefficients $\ln {\cal D}^{{\rm (int)} ab\rightarrow
cd}_{I\, N}$ whose NLL expansion reads: 
\beeq \label{lndeltams1}
\ln{\cal D}^{{\rm (int)} ab\rightarrow cd}_{I\,
N}(\as(\mu^2),Q^2/\mu^2) &\!\!=\!\!& \frac{D_{I\, ab \to c
d}^{(1)}}{2\pi b_0} \;\ln (1-2\lambda) + {\cal O}\left(\as(\as \ln
N)^k\right) \, . 
\eeeq 

At this point it is worth mentioning that one expects the jet cross section defined by a cone algorithm to be sensitive also to radiation in a limited part of phase space. Such radiation gives rise to \emph{non-global} logarithms~\cite{Salam1}. For instance, secondary emissions coherently radiated into the jet cone from a large-angle soft-gluon outside the jet cone originates non-global terms. Such contributions appears first at next-to-next-to-leading order although may produce threshold logarithms at NLL \cite{Salam2}. As performed in the unpolarized case \cite{resjetunp}, we neglect those contributions in our approximation.      

In order to obtain a resummed cross section in $x_T^2$ space, one needs 
an inverse Mellin transform. Here one has to deal with the singularity
in the perturbative strong coupling constant in 
Eqs.~(\ref{Dfct})-(\ref{Dintfct}), which manifests itself also in the 
singularities of the functions $h^{(1,2)}$ and $f^{(1,2)}$ above at 
$\lambda=1/2$ and $\lambda=1$. We implement the Minimal Prescription~\cite{Catani:1996yz}, which relies on the use of the 
NLL expanded forms Eqs.~(\ref{lndeltams})-(\ref{lndeltams1}), 
and on choosing a Mellin contour in complex-$N$ space that 
lies to the {\it left} of the poles at $\lambda=1/2$ and $\lambda=1$ 
in the Mellin integrand:
\begin{align}
\label{hadnmin}
\f{p_T^3\, d\Delta \sigma^{\rm (res)}(x_T)}{dp_T} &=
\;\int_{C_{MP}-i\infty}^{C_{MP}+i\infty}
\;\frac{dN}{2\pi i} \;\left( x_T^2 \right)^{-N}
\Delta \sigma^{\rm (res)}(N) \; ,
\end{align}
where $b_0\as(\mu^2)\ln C_{MP}<1/2$, but all other poles
in the integrand are as usual to the left of the contour.

Finally, in order to make full use of the available fixed-order 
cross section~\cite{nlojet}, which in our case is NLO (${\cal O}(\as^3)$), 
we match the resummed cross section to the NLO one. 
We expand the resummed cross section to ${\cal O}(\as^3)$, subtract the 
expanded result from the resummed one, and add the full NLO cross section:
\begin{align}
\label{hadnres}
\f{p_T^3\, d\Delta \sigma^{\rm (match)}(x_T)}{dp_T} &= \sum_{a,b,c}\,
\;\int_{C_{MP}-i\infty}^{C_{MP}+i\infty}
\;\frac{dN}{2\pi i} \;\left( x_T^2 \right)^{-N+1}
\; \Delta f_{a/{H_1}}(N,\mu^2) \; \Delta f_{b/H_{2}}(N,\mu^2) \;
 \nn \\
&\times \left[ \;
\Delta \hat{\sigma}^{\rm (res)}_{ab\to cd} (N)
- \left. \Delta \hat{\sigma}^{{\rm (res)}}_{ab\to cd} (N)
\right|_{{\cal O}(\as^3)} \, \right]
+\f{p_T^3\, d\Delta \sigma^{\rm (NLO)}(x_T)}{dp_T}
 \;\;,
\end{align}
where $\Delta \hat{\sigma}^{{\rm (res)}}_{ab\to cd} (N)$ is the 
polarized resummed cross section for the partonic channel $ab\to cd$ 
as given in Eq.~(\ref{eq:res}). In this way, NLO is taken into account 
in full, and the soft-gluon contributions beyond NLO are resummed to NLL. 
Any double-counting of perturbative orders is avoided.


\section{Phenomenological Results \label{sec4}}
We are now in the position to present numerical results for the 
threshold-resummed spin-dependent cross section and spin asymmetry in 
single-inclusive jet production in $pp$ collisions. We will focus 
here on $pp \to {\rm Jet} \, X$ at $\sqrt{S}=200$~GeV 
and at $\sqrt{S}=500$~GeV at the RHIC collider.

For our calculations, we need to select a set of parton 
distributions. To study the 
sensitivity of the measured spin asymmetries on the spin-dependent
parton densities, in particular the gluon density 
$\Delta g$, we mainly rely on the ``de Florian--Sassot--Stratmann--Vogelsang (DSSV)''~\cite{dssv} and the 
``Gl\"{u}ck--Reya--Stratmann--Vogelsang''~\cite{grsv}
distributions. The GRSV offers various sets, distinguished 
mostly by the size of $\Delta g$. 
The set labeled ``GRSV'' corresponds to the standard set. To compute the spin asymmetry we will
also use the ``de Florian--Sassot''~\cite{DS}  ``DSiii+'' and the GRSV ``max G'' set with a much larger polarized gluon content. 
Although sets with a rather large gluon polarization have been ruled out by the available data~\cite{dssv}, we will use them as a way to analyze the impact of the resummation in scenarios with the polarized cross section dominated by the gluon initiated subprocess.
 For the spin-averaged cross section, we employ the CTEQ6M~\cite{cteq} set throughout.

\begin {figure}[t]
\begin{center}
\includegraphics[width = 3.0in]{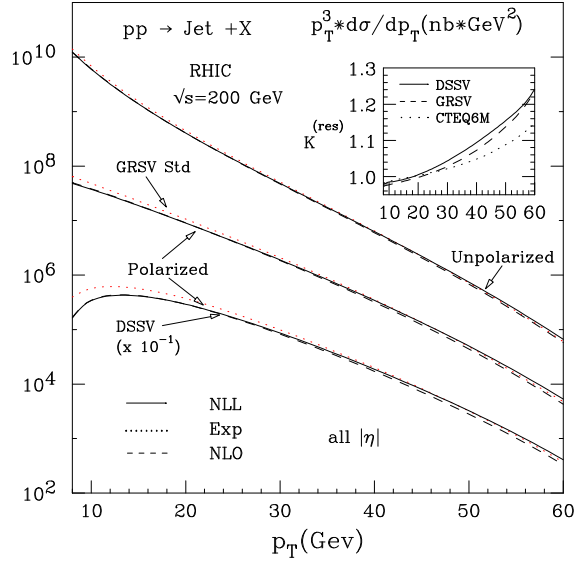}  
\end{center}
\caption{{\it NLO and NLL resummed cross sections for polarized 
$pp \to \rm{Jet}\, X $ at $\sqrt{S}=200$~GeV, for various sets of 
spin-dependent parton distributions~\cite{dssv,grsv}. We also
show the ${\cal O}(\as^3)$ expansions of the resummed cross sections,
and the analogous results in the unpolarized case.  For better 
visibility, we have applied numerical factors to some results, 
as indicated in the figure. In the upper right inset, we present the 
ratios between the NLL resummed cross section and the NLO one.}
\label{figres200} }
\end{figure}

In Fig.~\ref{figres200} we present our results for both the spin-averaged and spin-dependent
cross sections at $\sqrt{S}= 200$~GeV, integrated over all jet 
pseudorapidities $\eta$. We set all scales as $\mu_R=\mu_F=p_T$ and the jet size cone $R=0.4$. The full NLO cross
section based on the calculation in~\cite{nlojet}, as well as 
the NLL resummed predictions are shown. We also display the expansions of the
resummed cross section to ${\cal O}(\as^3)$, which is the first order
beyond LO. As can be observed at large transverse momentum, NLO, resummed cross section to ${\cal O}(\as^3)$ and NLL are practically indistinguishable. 
Towards lower $p_T\lesssim 35$ GeV, the expansions slightly 
overestimate the NLO cross section, which is expected since one 
is further away from the threshold regime and therefore the
soft-gluon approximation tends to become less reliable. The effect is more noticeable in the polarized case, and particularly for the DSSV set, because of the reduced impact of the gluon initiated processes (see the discussion below).  

\begin {figure}[h!]
\begin{center}
\includegraphics[width = 4.5in]{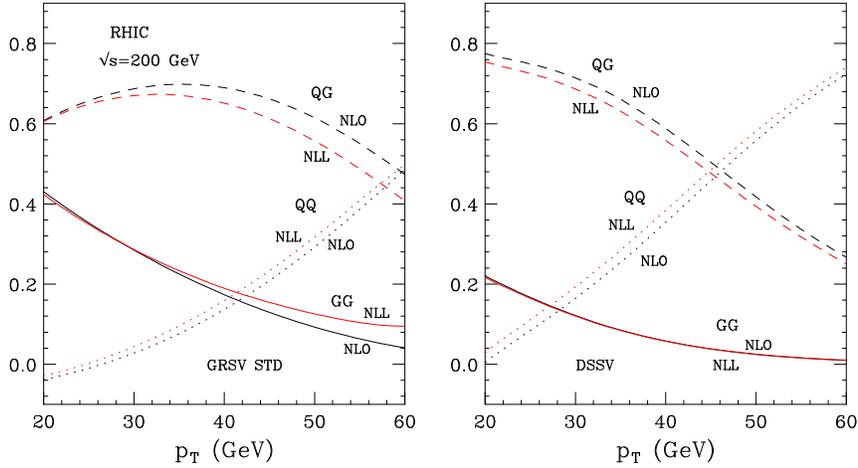}  
\end{center}
\caption{{\it NLO and NLL relative contributions of the different subprocesses to the single-inclusive jet cross sections in polarized $pp$ collisions at $\sqrt{S}=200$~GeV and jet cone size $R=0.4$}.
\label{partonic200}}
\end{figure}

The inset in Fig.~\ref{figres200} shows the resummed ``$K$-factors''
for the cross sections, defined as the ratios of the
resummed to the NLO cross sections (polarized or unpolarized) :
\begin{equation}
\label{eq:kres}
K^{{\rm (res)}} = \f{{d\sigma^{\rm (match)}}/{dp_T}}
{{d\sigma^{\rm (NLO)}}/{dp_T}}\, .
\end{equation}
As can be observed, $K^{{\rm (res)}}$ is larger than $1$ and grows with the transverse momentum of the jet, meaning that the resummation 
results in a overall enhancement over NLO. 
It is interesting to notice that the polarized $K^{{\rm (res)}}$ is a bit larger, for both sets of
spin-dependent parton distributions, than the one for the  spin-averaged case. This immediately
implies that the spin asymmetry $A_{LL}^{Jet}$ will be slightly increased when NLL effects are taken into account, at variance with what we found for single-inclusive hadron production $pp\to \pi^0 X$ in~\cite{dwf,dwf2}.


In order to study the role of each partonic channel and the differences originated by the use of the DSSV and GRSV pdf sets, we plot in Figure~\ref{partonic200} the relative contributions to the polarized cross section from the different initial parton combinations. We have grouped together $qq', q\bar{q}', qq$ and $q\bar{q}$ labelled as $QQ$. As expected, for a set with a smaller gluon polarization as DSSV, the impact of the pure quark contribution 
is larger than for GRSV. The variation in the corresponding partonic ratios due to the inclusion of threshold effects is rather moderate, with a bigger impact in scenarios with large gluon polarization, which can be easily understood due to the larger color factor for initial state gluons.

\begin {figure}[h!]
\begin{center}
\includegraphics[width = 3.0in]{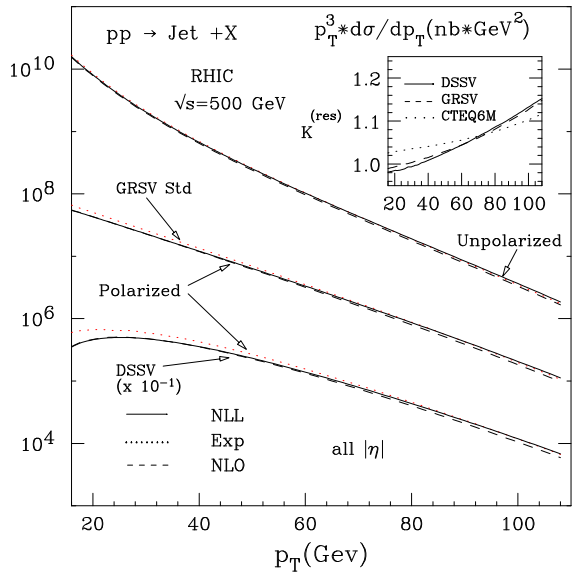}  
\end{center}
\caption{{\it Same as Fig.~\ref{figres200} but for $\sqrt{S}=500$ GeV and $R=0.7$.}
\label{figres500}}
\end{figure}

In Figure~\ref{figres500} we provide the corresponding phenomenological predictions for $p p \to \rm{Jet}\, X$ at $\sqrt{S}=500$ GeV, also relevant for the spin physics program at RHIC. Here we have chosen $R=0.7$.  From the inset of Fig.~\ref{figres500} one can observe that the resummation effects 
are generally smaller at $\sqrt{S}= 500$~GeV than at $\sqrt{S}=200$~GeV as expected from the fact that one is further away from threshold here. The relative  partonic contributions at $\sqrt{S}= 500$~GeV are very similar to those at $\sqrt{S}= 200$~GeV. We have explicitly checked that for both energy scenarios studied and the values of transverse momentum analyzed in this paper, the SCA agrees with the full NLO calculation at a percent level. 

\begin {figure}[h!]
\begin{center}
\includegraphics[width = 7.0in]{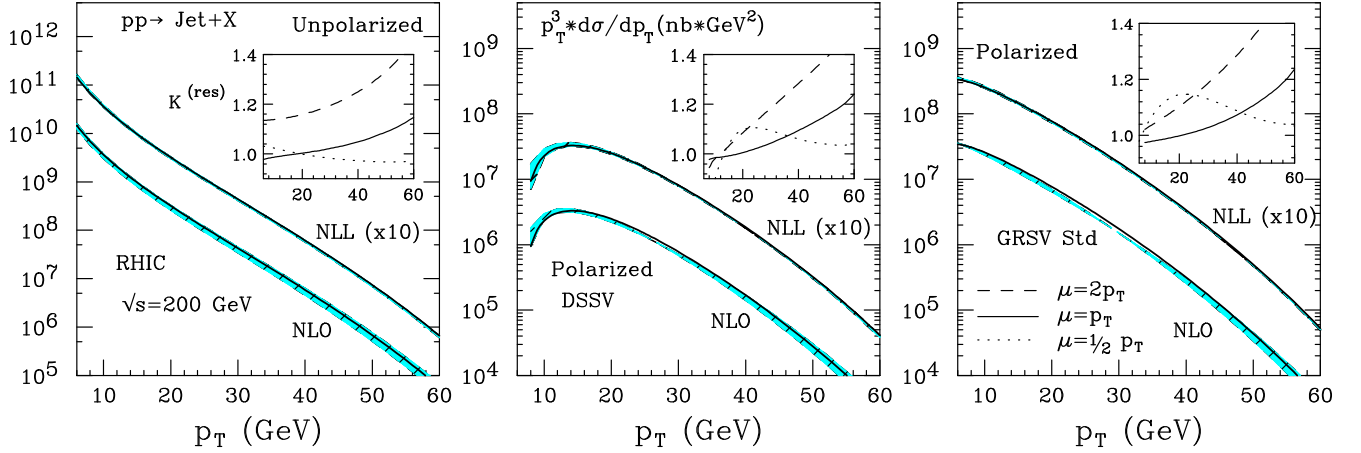}  
\end{center}
\caption{{\it NLO and NLL $ pp \to \rm{Jet}\, X $ cross sections at $\sqrt{S}=200$~GeV for different scales  $\mu_R=\mu_F=\zeta p_T$ with $\zeta =1/2, 1, 2$. The unpolarized cross section is shown on the left whereas the results for the polarized case are analyzed for DSSV (center) and GRSV std (right). }
\label{NLL200}}
\end{figure}

\begin {figure}[h!]
\begin{center}
\includegraphics[width = 7.0in]{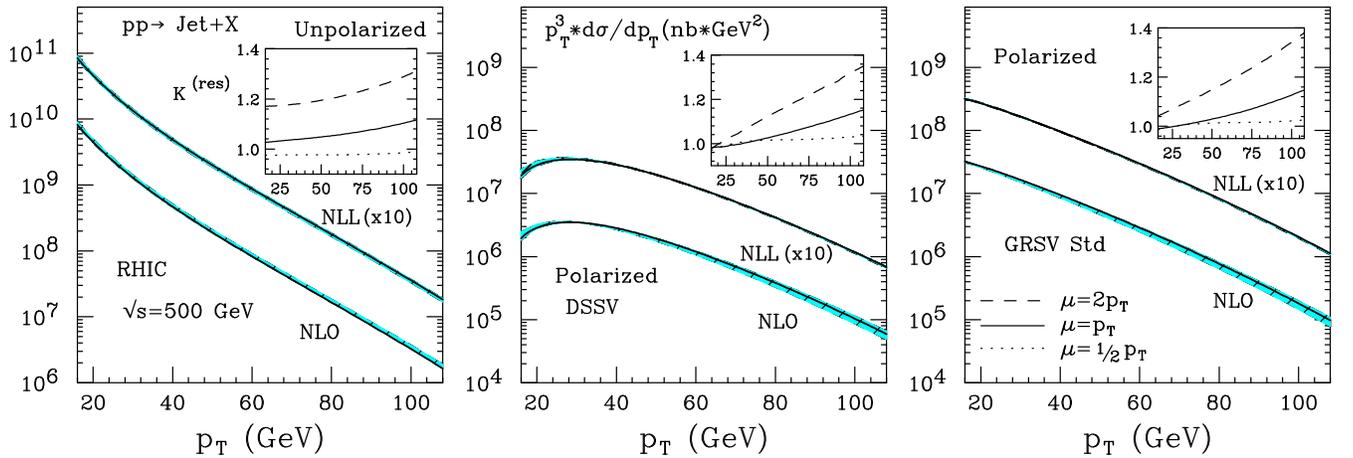}  
\end{center}
\caption{{\it Same as figure \ref{NLL200}, but for $\sqrt{S}=500$~GeV. }
\label{NLL500}}
\end{figure}

After having established the importance of the threshold corrections for jet production in polarized hadronic collisions at the default scales $\mu\equiv \mu_R=\mu_F=p_T$, we concentrate on the impact of the resummation over the factorization and renormalization scale dependence. 
Figure~\ref{NLL200} shows the NLO and NLL cross section at $\sqrt{S}= 200$~GeV for the choice of scales $\mu =\zeta p_T$ with $\zeta =1/2, 1, 2$. 
The most noticeable effect is a remarkable reduction in the scale dependence for both unpolarized and polarized cross sections when the threshold effects are resummed to NLL accuracy. For example, in the polarized case at $p_T=40$ GeV the width of the scale dependence band is reduced from $9 \%$ at NLO up to $1\%$  at NLL.
In the upper insets it can be observed the $K$-factors corresponding to the different scales. The corresponding results for $pp\to \rm{Jet} \, X$ at $\sqrt{S}= 500$~GeV are shown in Figure~\ref{NLL500}, finding a similar trend concerning the reduction of the scale dependence.

\begin {figure}[h!]
\begin{center}
\includegraphics[width = 5.0in]{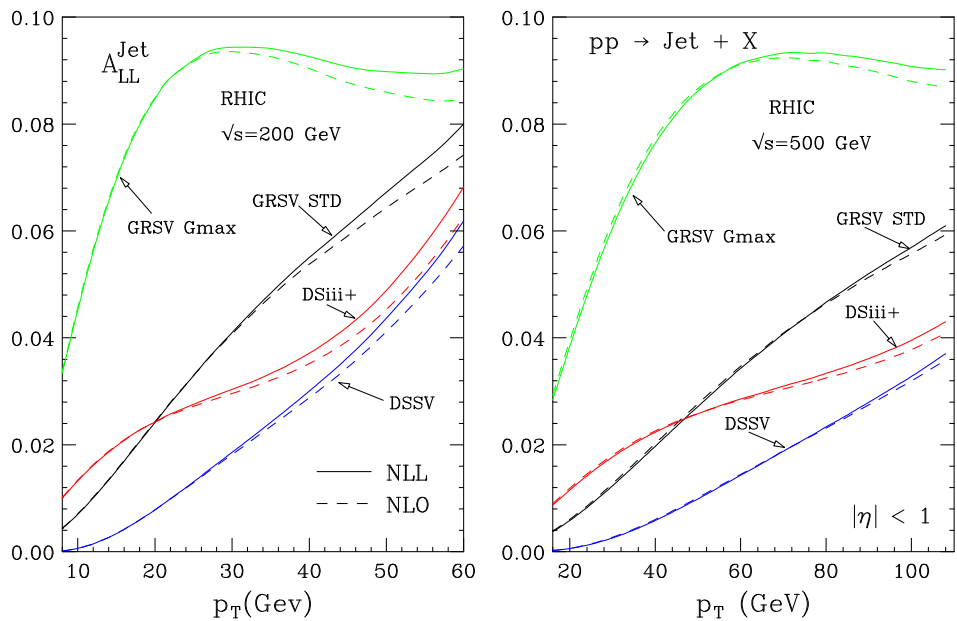}  
\end{center}
\caption{{\it Results for the double-spin asymmetry $A^{Jet}_{LL}$ 
at NLO (dashes) and NLL (solid) for various sets of polarized
parton distributions, at $\sqrt{S}=200$~GeV (left) and $\sqrt{S}=500$~GeV (right).}
\label{fig5}}
\end{figure}

Finally we present  in Figure~\ref{fig5} our phenomenological predictions for the spin asymmetry $A_{LL}^{Jet}$ at $\sqrt{S}=200$~GeV and at $\sqrt{S}=500$~GeV. In order to match the experimental conditions of the STAR collaboration, covering only a finite region in rapidity $|\eta|\le 1$, we multiply the $K-$factors defined in Eq.(\ref{eq:kres}) by the NLO cross section computed in the experimentally accessed rapidity regime.
For the sake of simplicity we have set both factorization and renormalization scales to the default value. As we have mentioned, the sets GRSV ``G-max'' and ``DSiii+'' were also used in order to study $A_{LL}^{Jet}$  for different parton distribution functions. As can be observed in Fig.~\ref{fig5} the NLO and NLL results are similar up to $p_T \approx 30$~GeV in the $\sqrt{S}=200$~GeV case and $p_T \approx 80$~GeV in the $\sqrt{S}=500$~GeV case. For higher $p_T$ values we observe a small increase of $A_{LL}^{Jet}$ when going from NLO to NLL, as predicted above. This is more noticeable in the case of parton distributions with a larger gluon polarization, like the  GRSV ``G-max'' set.
From this findings one can conclude that the effect of the resummation of large logarithmic corrections for single-inclusive jet production in polarized $pp$ collisions is rather modest at the level of the asymmetry and can be omitted in the analysis of the experimental data.  

\section{Conclusions  \label{sec5}}

We have studied in this paper the NLL resummation of threshold
logarithms in the partonic cross sections relevant for the process
$pp\to \rm{Jet}\, X$, where the initial protons are longitudinally polarized and jets are produced with a high transverse momentum. 
We find that for the spin-dependent case the resummation effects increase the cross section although in a rather modest way. The main effect of the resummation consists in a considerable reduction on the scale dependence resulting, at the same time, in a stabilization and confirmation of the fixed order predictions.
We have analyzed the impact of the resummation in the spin asymmetry $A_{LL}^{Jet}$ in the cases of  $\sqrt{S}= 200$~GeV and  $\sqrt{S}= 500$~GeV relevant for RHIC, finding that the modification due to the NLL corrections can be neglected in a first analysis of the data.  

\section*{Acknowledgments}
We are grateful to Werner Vogelsang for helpful discussions. The work of D.deF. has been partially supported by Conicet,
 UBACyT, ANPCyT and the Guggenheim Foundation.
 The work of F.W. has been supported by UBACyT. 

\newpage

\appendix

\section*{Appendix }

The spin-dependent Born cross sections and $\Delta G_{I\, ab\to cd}$ coefficients are the same as for hadron production and can be found in the appendix A of Ref.~\cite{dwf}.  As mentioned above, the coefficients $D_{I\, ab\to cd}$ are the same 
as for the unpolarized case; they are given in Ref.~\cite{DW1}. 
Only the $\Delta C_{ab\to cd}^{(1)}$ coefficients need to be calculated. 
Since the $\Delta C_{ab\to cd}^{(1)}$  have rather lengthy expressions,
we only give their numerical values for $N_f=5$ and the
factorization and renormalization scales set to $\mu=Q$. 

\begin{description}
\item[$qq'\to Jet \, X$:]
\begin{equation}
\Delta C^{(1)}_{1\, qq'\to Jet \, X} = 15.5934 + 1.38763 \,\, \log(\frac{R}{2}).
\end{equation}
\item[$q\bar{q'}\to Jet \, X$:]
\begin{equation}
\Delta C^{(1)}_{1\, q\bar{q'}\to Jet \, X}= 18.3644 + 1.38763 \,\, \log(\frac{R}{2}).
\end{equation}
\item[$q\bar{q}\to Jet \, X$:]
\begin{equation}
\Delta C^{(1)}_{1\, q\bar{q}\to Jet \, X}= -29.5367 + 4.18905 \,\, \log(\frac{R}{2})  . 
\end{equation}
\item[$qq\to Jet \, X$:]
\begin{equation}
\Delta C^{(1)}_{1\, qq\to Jet \, X}= 12.1987 + 1.38763 \,\, \log(\frac{R}{2}).
\end{equation}
\item[$qg\to Jet \, X$:]
\begin{equation}
\Delta C^{(1)}_{1\, qg\to Jet \, X}=13.2364 + 2.58824 \,\, \log(\frac{R}{2}).
\end{equation}
\item[$gg\to Jet \, X$:]
\begin{equation}
\Delta C^{(1)}_{1\, gg\to Jet \, X}= 13.6432 + 3.94682 \,\, \log(\frac{R}{2}).
\end{equation}
\end{description}

In the  expressions above, $R$ is the Jet cone size.


\end{document}